%% file: main.tex
\documentclass[review]{cas-dc}
\usepackage{xcolor}
\usepackage{graphicx}
\usepackage{rotating}
\usepackage{enumitem}
\usepackage{amsmath,amsfonts}
\usepackage[numbers]{natbib}

\newcommand{\Z}{\Omega}

\newcommand{\x}{\bm{x}}
\newcommand{\C}{C(\x,t)}
\renewcommand{\u}{\bm{u}}
\newcommand{\ut}{\u(\x,t)}

\newcommand{\R}{\mathbb{R}}

\usepackage{algorithm}
\usepackage{algpseudocode}
\definecolor{citeblue}{RGB}{0,102,184}
\hypersetup{citecolor=citeblue, urlcolor=citeblue}
\usepackage{float}
\usepackage{bm}
\allowdisplaybreaks

\begin{document}
\let\WriteBookmarks\relax
\def\floatpagepagefraction{1}
\def\textpagefraction{.001}
\newcommand{\done}[1]{\textcolor{black}{#1}}
\newcommand{\seconddone}[1]{\textcolor{black}{#1}}
\definecolor{donecolor}{named}{black}

\shorttitle{Operator learning for energy-efficient building ventilation control}

\shortauthors{Yuexin Bian et~al.}

\title [mode = title]{Operator learning for energy-efficient building ventilation control with computational fluid dynamics simulation of a real-world classroom} 



\author[1]{Yuexin Bian}
\affiliation{{Department of Electrical and Computer Engineering, University of California San Diego},
    addressline={}, 
    city={La Jolla},
    postcode={92037}, 
    country={USA}}
\credit{Investigation, Methodology, Software, Writing – original draft, Writing - review \& editing}

\author[2]{Oliver Schmidt}
\affiliation{{Department of Mechanical and Aerospace Engineering, University of California San Diego},
    addressline={}, 
    city={La Jolla},
    postcode={92037}, 
    country={USA}}
\credit{Investigation, Methodology, Software, Writing - review \& editing}

\author[1]{Yuanyuan Shi}[
orcid=0000-0002-6182-7664
]
\credit{Supervision, Investigation, Methodology, Writing - review \& editing}
\cormark[1]

\cortext[cor1]{Corresponding author}



\begin{abstract}
Energy-efficient ventilation control plays a vital role in reducing building energy consumption while ensuring occupant health and comfort. While Computational Fluid Dynamics (CFD) simulations provide detailed and physically accurate representation of indoor airflow, their high computational cost limits their use in real-time building control.
In this work, we present a neural operator learning framework that combines the physical accuracy of CFD with the computational efficiency of machine learning to enable building ventilation control with the high-fidelity fluid dynamics models. Our method jointly optimizes the airflow supply rates and vent angles to reduce energy use and adhere to air quality constraints. We train an ensemble of neural operator transformer models to learn the mapping from building control actions to airflow fields using high-resolution CFD data. This learned neural operator is then embedded in an optimization-based control framework for building ventilation control.
Experimental results show that our approach achieves significant energy savings compared to maximum airflow rate control, rule-based control, as well as data-driven control methods using spatially averaged CO$_2$ prediction and deep learning–based reduced-order model, while consistently maintaining safe indoor air quality.
These results highlight the practicality and scalability of our method in maintaining energy efficiency and indoor air quality in real-world buildings.
\end{abstract}




\maketitle

\section{Introduction}
\input{intro}

\section{Problem formulation}
\input{problem2}

\section{The BEAR-CFD data}\label{sec:data}
\input{CFD_simulation}

\section{Methodology}
\input{method}

\section{Numerical experiments}
\input{experiment}

\section{Conclusion and future work}
In this work, we propose a novel operator learning framework for energy-efficient ventilation control. Our approach involves training an ensemble of neural operator transformer models to learn the mapping from past CO$_2$ fields, occupancy levels, and control actions to future CO$_2$ fields. The ensemble model demonstrates superior predictive performance compared to individual models. We further integrate the learned neural operator into building ventilation optimization to optimize control actions. Using CFD simulations, we validate the proposed approach achieves substantial energy savings compared to maximum airflow control, rule-based control, and data-driven methods based on average CO$_2$ predictions and reduced-order modeling. In addition, compared to baseline control, our method maintains similar energy consumption while significantly reducing CO$_2$ violations. We open-source the CFD data to facilitate further research in developing machine learning models for building ventilation control with PDE-based models.


Promising future directions include extending the framework to more complex building environments, such as multi-zone systems. In addition, real-world experiments with integrated sensing and actuation systems are planned to validate the framework’s practical performance. 

\section*{Acknowledgments}
The authors gratefully acknowledge Ling Zhong for the framework visualization design. This work is supported by a Schmidt Sciences AI2050 Early Career Fellowship, NSF grant ECCS-2442689, and DOE grant DE-SC0025495.

\bibliographystyle{elsarticle-num}
\bibliography{refs}

\appendix
\input{appendix}

\end{document}

%% file: intro.tex
\color{donecolor}
Buildings account for nearly 40\% of global energy consumption~\cite{chua2013achieving}, with Heating, Ventilation, and Air Conditioning (HVAC) systems being among the primary contributors. Effective ventilation is critical not only for reducing energy use but also for maintaining indoor air quality, which directly impacts occupant health and comfort~\cite{chenari2016towards}. The urgency of this issue has grown in light of the COVID-19 pandemic, as building system operators and public health agencies have increasingly emphasized the need for adaptive ventilation systems that respond to occupancy and pollutant levels throughout the indoor space~\cite{hosseinloo2023data}.

Despite this urgency, most building HVAC systems still rely on fixed or rule-based ventilation strategies~\cite{shi2025event, REHAV, bian2023bear}, which are often overly conservative. For instance, maximum fresh-air intake policies implemented at UC San Diego~\cite{bian2023bear} in response to health concerns have led to 2–2.5$\times$ increases in energy usage. 
This highlights the critical trade-off between maintaining indoor air quality and minimizing energy consumption, and the need for intelligent control methods that balance these competing demands.

Recent advances in data-driven control approaches, including model predictive control (MPC)~\cite{drgovna2020all, chen2018optimal, gao2023energy, su2023maintaining} and reinforcement learning (RL)~\cite{du2021intelligent, yang2021towards, wang2024energy, shang2023developing, li2023proactive}, have shown strong potential for building ventilation management. However, most existing methods represent indoor air states—such as CO$_2$ concentration or temperature—using spatially aggregated variables, typically measured at a single point or averaged over an entire zone.
For example, MPC and RL-based strategies often model CO$_2$ dynamics using ordinary differential equations (ODEs)~\cite{shang2023developing, li2022tube} or neural networks~\cite{drgovna2020all,chen2018optimal,li2023proactive} that predict future mean values based on current measurements and control inputs. While computationally efficient, these low-dimensional representations ignore spatial variations of indoor airflow velocity fields and pollutant distribution that arise from the ventilation layout, control actions, and occupancy patterns. As a result, such controllers fail to capture the localized effects of ventilation decisions. They may over-ventilate the entire space to maintain air quality at a single location, wasting energy, or overlook under-ventilated regions that compromise occupant health and comfort.
This gap highlights the need for \emph{spatiotemporal} indoor airflow modeling to accurately capture how ventilation control actions affect air quality—crucial for maintaining occupant health while enabling energy-efficient building HVAC operation.

To accurately model airflow dynamics and indoor air quality, Computational Fluid Dynamics (CFD) simulations are widely adopted in building ventilation research~\cite{tian2018building}. For example, Bianco et al.~\cite{bianco2023cfd} employed CFD to assist in the design of ventilation units for buildings, while Gao et al.~\cite{gao2024rapid} used CFD to analyze airflow fields in an isothermal chamber under fixed ventilation rates. 
Bulinska et al.~\cite{bulinska2014experimentally} modeled CO$_2$ distribution in a bedroom to inform optimal sensor placement, while Mou et al.~\cite{mou2022computational} simulated airflow and CO$_2$ concentration in a seminar room, capturing complex 3D ventilation patterns. Similarly, Ning et al.~\cite{ning2016computational} used CFD to evaluate the influence of supply outlet height on indoor air distribution in a bedroom. Extending to classroom environments, Mahmoud~\cite{mahmoud2024numerical} applied CFD to analyze the dispersion of human-generated aerosols and CO$_2$, offering insights into ventilation design for indoor air quality.

\begin{figure*}[t]
    \centering
    \includegraphics[width=0.85\linewidth]{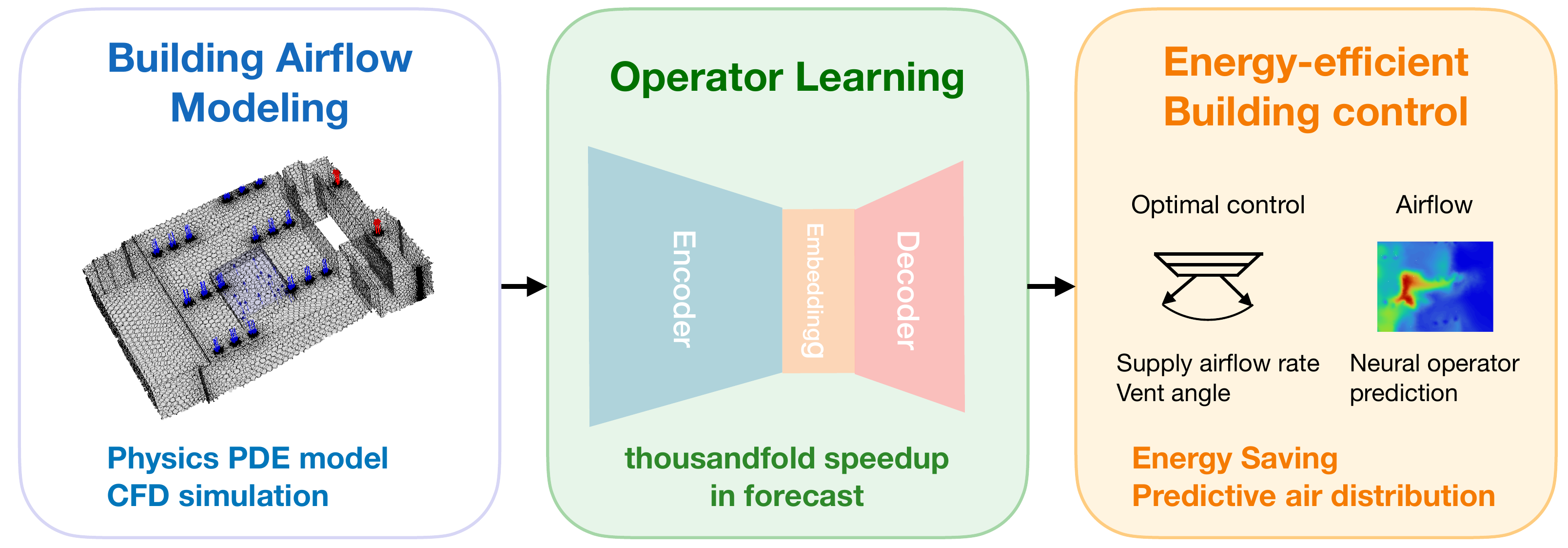}
    \caption{Schematic of our data-driven operator learning framework for energy-efficient ventilation control. Computational fluid dynamics (CFD) simulations are used to model complex 3D airflow and CO$_2$ spatiotemporal dynamics. An ensemble of neural operator transformer models is trained to learn the mapping between ventilation control actions and airflow field evolution. Leveraging high-fidelity simulation data, our approach enables real-time optimization of ventilation strategies to minimize energy consumption while maintaining indoor air quality.
    }
    \label{fig:framework1}
\end{figure*}

Although CFD simulations can accurately model indoor airflow fields, solving the governing partial differential equations (PDEs) using numerical solvers, such as finite element and finite volume methods, are computationally expensive~\cite{michoski2020solving}. This high cost makes it impractical to embed CFD solvers directly into real-time control loops, where rapid decisions must be made in response to changing occupancy and environmental conditions. Consequently, most CFD-based studies~\cite{mou2022computational, ning2016computational, mahmoud2024numerical} focus on offline analysis and system design rather than real-time control. Our prior work~\cite{bian2024ventilation} attempted to bridge this gap using a differentiable PDE-based framework that optimizes control via adjoint methods. While this approach enabled physically grounded optimization, it required repeatedly solving PDE-constrained optimization and was limited to a 2D setting with simplified geometry, highlighting a key scalability bottleneck.

To address these limitations, in this work, we propose a data-driven operator learning framework to approximate the input–output behavior of CFD simulations. Neural operators are designed to learn mappings between \emph{infinite-dimensional} function spaces, enabling them to approximate the solution operator of PDEs across a family of initial and boundary conditions~\cite{li2021fourier,lu2021learning}.
Previous studies~\cite{zhou2021neural,warey2020data} have used conventional neural networks to learn indoor air dynamics from CFD data. However, these models operate in finite-dimensional spaces and typically require fixed spatial grids, making them inflexible and often requiring retraining under new control or occupancy settings.
In contrast, neural operators learn mappings between continuous functions, allowing predictions to be queried at arbitrary spatial locations within the room. This functional formulation captures how ventilation control actions shape the entire indoor airflow and air quality distribution, enabling flexible and efficient ventilation control across diverse operating conditions.
Several neural operator architectures have demonstrated strong performance in modeling complex physical systems. For example, Deep Operator Networks (DeepONets) \cite{lu2021learning} and Fourier Neural Operators (FNOs) \cite{li2021fourier} enable efficient approximation of fluid dynamics with significantly reduced computational cost. These models have shown success in applications such as turbulent flow prediction \cite{wang2024prediction}, vehicle aerodynamics~\cite{li2024geometry}, and indoor airflow modeling~\cite{gao2024rapid}. More recently, General Neural Operator Transformers (GNOTs)~\cite{hao2023gnot} have extended these capabilities to irregular spatial domains and multiple input functions, making them well suited for modeling building airflow with complex geometries and diverse inputs (e.g., control actions, occupancy, and past room states). 

\color{black}

\subsection{Summary of contributions}
In this work, we propose a data-driven operator learning framework for energy-efficient building ventilation control, validated in a real-world classroom using CFD simulations (Figure~1). While prior studies~\cite{gao2024rapid, ding2025fourier} have applied neural operators for airflow prediction, our work advances this line in two key ways: (i) we propose an ensemble neural operator transformer model that enhances prediction accuracy and robustness, and (ii) we introduce the integration of operator learning into closed-loop control for building energy management. This is substantially more challenging, as performance must account for both predictive accuracy and reliable control under energy and air quality constraints.

We benchmark our approach against data-driven ventilation control methods using averaged models (e.g., MLPs) and deep learning–based reduced-order models (e.g., U-Nets)~\cite{pant2021deep}. Unlike MLPs, neural operators capture full spatial airflow fields, and unlike reduced-order models, they learn mappings between function spaces rather than pointwise values. This functional representation captures how control actions shape the entire airflow patterns and resulting indoor air quality. CFD-validated experiments demonstrate that these advantages enable neural operators to achieve more energy-efficient and healthier indoor environments compared to finite-dimensional alternatives. Our key contributions are summarized as follows:



\begin{itemize}
\item \seconddone{\emph{Ensemble neural operator transformer:} We propose an ensemble neural operator transformer model to predict building airflow velocity and CO$_2$ fields under varying HVAC control actions and occupancy levels.
This model improves prediction accuracy compared to individual neural operator models and achieves a remarkable speed-up of over 250,000× relative to CFD simulations, enabling accurate indoor air quality prediction with high computational efficiency.}
\item \seconddone{\emph{Building control with operator learning:} We incorporate the learned neural operator into an optimization framework for energy-efficient ventilation control subject to air quality constraints. Compared to maximum airflow control, rule-based control, and data-driven control using averaged prediction or reduced-order models, our approach enables more effective closed-loop building ventilation control under varying occupancy, achieving lower energy use and significantly fewer CO$_2$ violations.}
\item \seconddone{\emph{Opensource building CFD dataset:} We develop and release a high-fidelity, open-access building dataset derived from CFD simulations of a real-world classroom. Along with the dataset, we publish the underlying 3D room model and the files required to reproduce and extend the simulations. The dataset includes airflow velocity fields and CO$_2$ fields under diverse HVAC control actions and occupancy levels, providing a reproducible benchmark for training and evaluating machine learning models in building ventilation control applications.}
\end{itemize}

%% file: problem2.tex
\begin{figure*}
    \centering
    \includegraphics[width=0.85\linewidth]{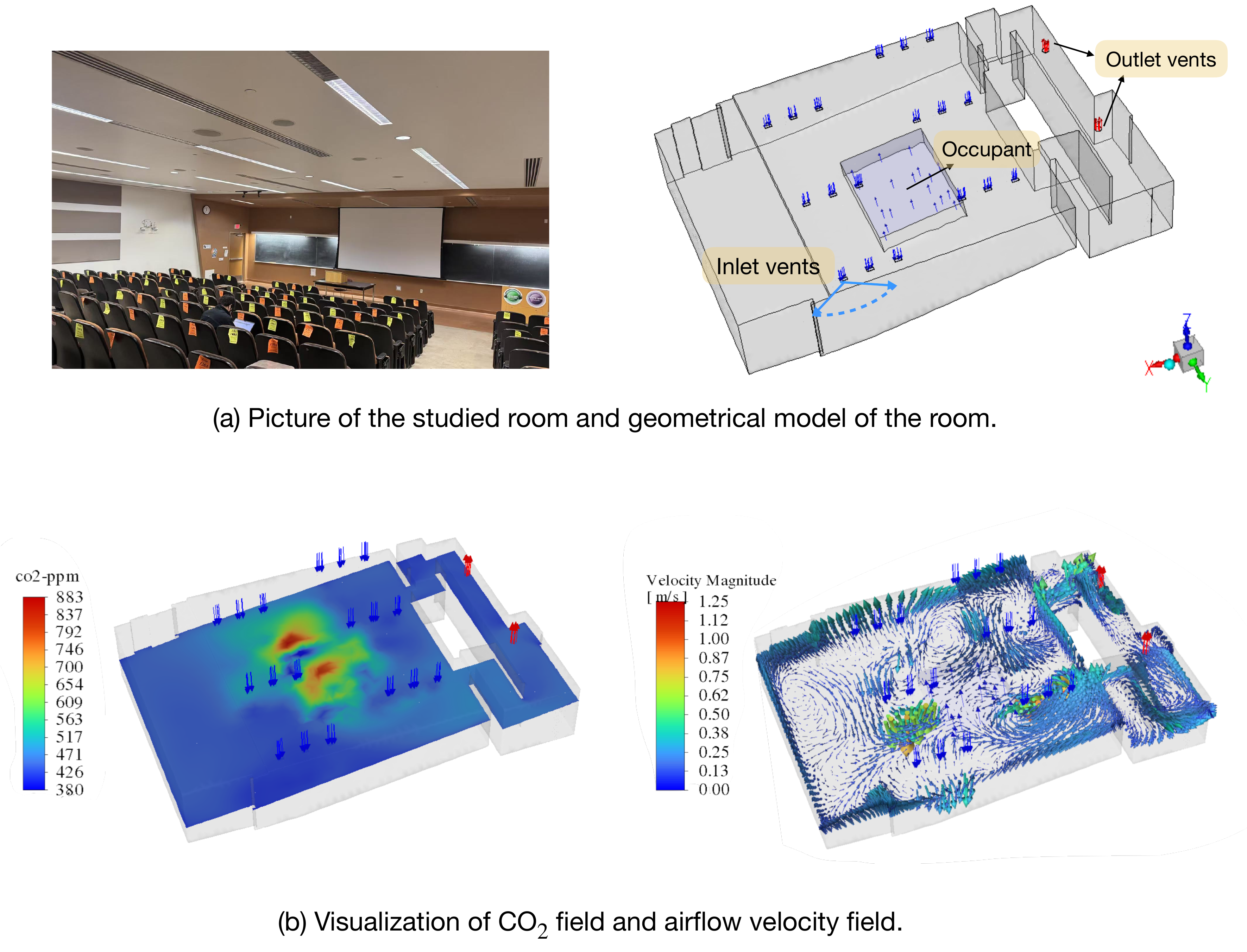}
    \caption{(a) The picture of the studied room and the geometry of the CFD model: a classroom with a ventilation system including 18 inlet vents and 2 outlet vents on the ceiling. \done{Occupants are modeled as a single rectangular cuboid with a prescribed CO$_2$ mass flux to represent CO$_2$ effects of varying occupancy levels.} (b) Visualization of the CFD simulation results of CO$_2$ concentration and airflow velocity fields - one example from the developed CFD dataset. }
    \label{fig:cadmodel}
\end{figure*}
In this section, we introduce the governing dynamics of indoor airflow and CO$_2$ transport, and formulate the learning and control problems. The CFD simulation setup and dataset description are provided in Section~\ref{sec:data}.

\subsection{Governing equations for CO$_2$ dynamics}
Let $\Omega \subset \R^3$ be the spatial domain of interests, and let $t \in \R^+$ denote time. We define $C(\x, t)$ as the CO$_2$ concentration at spatial location $\x \in \Omega$ and time $t$. Let $m(t) = \begin{bmatrix}
    m^r(t)  & m^a(t) 
\end{bmatrix} \in \R^{12}$ be the control actions (airflow rates $m^r \in \R^6$ and airflow angles $m^a \in \R^6$) for the six groups of supply vents, and $n_p(t) \in \R$ denote the occupancy number. 
The distribution of CO$_2$ in an indoor environment follows the advection-diffusion equation~\cite{bian2024ventilation,bulinska2017cfd},
\begin{equation}
    \frac{\partial \C }{\partial t} + \ut \cdot \nabla \C = D_{\text{eff}} \nabla^2 \C + S(\x, t),
\end{equation}
where 
\begin{itemize}
    \item $\C$ is the CO$_2$ concentration (ppm) at location $\x$ and time step $t$.
    \item $\ut$ is the airflow velocity field obtained from CFD simulations, by solving the incompressible Navier-Stokes equations.
    The boundary conditions at the supply vents depend on the ventilation control $m(t)$, thereby allowing the control action to influence $\u(\x,t)$ throughout the domain. The numerical model for the airflow velocity field is presented in the Appendix-A.
    \item $D_{\text{eff}}$ is the diffusion coefficient for CO$_2$ in the air. 
    \item $S(\x,t)$ represents the CO$_2$ source term, which models occupant-generated CO$_2$ on the designated occupancy surface. This source depends on the occupancy level $n_p(t)$ and we assume that the exhaled air rate is 6 L/min per person following~\cite{he2022experimental} .
\end{itemize}

\subsection{Data-driven modeling with neural operator}
Solving PDEs numerically is computationally expensive, making traditional CFD models impractical for real-time building control applications.
To address this limitation, we aim to learn a neural operator $\mathcal{G}_\theta$ that efficiently maps historical CO$_2$ concentrations, control actions, and occupancy levels to future CO$_2$ concentration distributions.
Formally, the neural operator learns the following mapping,
\begin{equation}
    \mathcal{G}_\theta: \left(C(\x,\tau)_{\tau \in [t-H, t]}, m, n_p \right) \mapsto C(\x,\tau)_{\tau \in (t, t+T]}
\end{equation}
where $C(\x, \tau)_{\tau \in [t-H,t]}$ is the historical CO$_2$ fields over period $[t-H,t]$ and $C(\x, \tau)_{\tau \in (t, t+T]}$ is the predicted future CO$_2$ fields over the future interval $(t, t+T]$. Our forecasting approach incorporates historical CO$_2$ concentrations to account for temporal dependencies inherent to the system’s physics (e.g., diffusion and advection dynamics). We assume that the control $m$ and occupancy $n_p$ remain \emph{fixed} over $(t,t+T]$. 
In building control applications, the transient dynamics of airflow and CO$_2$ transport evolve rapidly over a short time horizon—typically within a few minutes. However, building control settings and occupancy levels generally remain constant during these intervals (e.g., a fixed HVAC setpoint over a 5-minute or 15-minute interval~\cite{bian2023bear}). This assumption thus maintains consistency between the physical model and practical control timescales.

In practice, we rarely have continuous access to the CO$_2$ distribution over space and time. Instead, we rely on a discretized approximation of the underlying functions. For simplicity and consistency, we reuse the symbols $t, H, T$ as discrete time indices. 
Let $\{x_i\}_{1 \leq i \leq N_x}$ be a spatial grid and $C_i^t \in \R$ be the CO$_2$ concentration at spatial grid point $x_i$ and time step $t$. We collect the recent history of CO$_2$ fields over the last $H$ time steps, along with the fixed parameters $m$ and $n_p$. Define the discrete input set
\begin{equation}
    \mathcal{A} = \{(x_i, C_i^{t-H:t})\}_{1 \leq i \leq N_x} \cup \{m\} \cup \{n_p\},
\end{equation}
although $\mathcal{A}$ is finite-dimensional, it represents the sampled version of the underying function $C(\x,\tau)$ over the history window $[t-H,t]$.
From this discrete representation, the neural operator $\mathcal{G}_\theta$ produces predicted future concentrations $\widehat{C}_{1\leq i \leq N_x}^{t+1:t+T}$(with short-hand notation $\widehat{C}$), which approximate the continuous output function $\widehat{C}(\x,\tau)$ for $\tau \in (t, t+T]$: 
\begin{equation}\label{eq:CO2predicton}
    \widehat{C}_{1\leq i \leq N_x}^{t+1:t+T} = \mathcal{G}_\theta (\mathcal{A}).
\end{equation}

\subsection{Ventilation control optimization}
The learned neural operator $\mathcal{G}_\theta$ can be integrated into the building ventilation control optimization problem. 
The resulting optimization problem is formulated in~\eqref{eq:problem}.
\begin{subequations}\label{eq:problem}
\begin{align}
    \min_{m = [m^r, m^a]} \, & w_1 \|\widehat{C}-C_{\text{target}}\|_2^2 + w_2 \|m-m^{(0)}\|_2^2 + w_3 \|m^r\|_1,  
    \label{eq:problem_obj}   \\
    \text{s.t.} \quad & 
    \widehat{C} \leftarrow \mathcal{G}_\theta (\{C_{1 \leq i \leq N_x}^{t-H:t}\} \cup \{m\} \cup \{n_p\}), \label{eq:problem_c1}  \\
    & \underline{m^r} \leq m^r \leq \overline{m^r}, \, \underline{m^a} \leq m^a \leq \overline{m^a}. \label{eq:problem_c2}
\end{align}
\end{subequations}
In the objective function~\eqref{eq:problem_obj}, the first term quantifies the deviation between predicted CO$_2$ concentrations $\widehat{C}$ and the desired CO$_2$ level $C_{\text{target}}$ over a prediction horizon $T$. This term is commonly used in indoor air quality ventilation control studies, as seen in prior works~\cite{li2022tube,sha2025online}.
The second term penalizes deviations of the optimized control actions from the previous control action $m^{(0)}$. 
This is important for real-world building management, where large deviations in control actions are typically avoided to maintain system stability and operational safety~\cite{ bian2024ventilation, zhang2021novel}. By encouraging control actions to remain close to $m^{(0)}$, we also reduce the risk of extrapolating beyond the model's training domain, thereby increasing the reliability of the predictions. 
The third term represents the energy consumption, measured through the L1 norm of the ventilation rate~\cite{bian2024ventilation}.  The coefficients $w_1, w_2, w_3$ balance the relative importance of these objectives.
Constraint~\eqref{eq:problem_c1} describes that the predicted CO$_2$ concentrations is obtained via the trained neural operator model in Eqn \eqref{eq:CO2predicton}, while constraint~\eqref{eq:problem_c2} ensures that both mechanical ventilation rates $m^r$ and vent angles $m^a$ remain within their physical limits.

\color{donecolor}
\paragraph{Remark 1.} In this study, we do not explicitly model or control indoor temperature in the problem formulation. This choice is motivated by the fact that CO$_2$ concentration responds on a much faster time scale than temperature, which evolves more slowly because of the building’s thermal inertia.
By focusing on the faster dynamics-airflow and CO$_2$ transport, we are able to
evaluate the proposed control strategy in a more responsive setting.
We note that thermal effects are still captured, as our CFD simulations solve the energy equation~\eqref{eq:energy_equation} which models the transport of thermal energy within the airflow and include occupant heat sources and boundary temperature conditions. 
Extending the formulation to include indoor temperature modeling and thermal control remains a valuable direction for future work, and our framework is compatible with such multi-objective settings.

\color{black}

%% file: CFD_simulation.tex
\color{donecolor}
The CFD simulation is developed based on a real-world classroom located in University of California, San Diego. The classroom is equipped with a ceiling-mounted ventilation system. 
The ventilation system includes 2 outlet vents and 18 inlet vents grouped into six zones, each allowing independent control of airflow rates and supply airflow angles to optimize energy efficiency while maintaining high indoor air quality.
Figure~\ref{fig:cadmodel}(a) illustrates the physical classroom and its corresponding 3D computer-aided design (CAD) model representation, and Figure~\ref{fig:cadmodel}(b) presents an example of the CO$_2$ concentration and velocity fields visualization. 
This section provides a detailed description of the simulation setup and the open-source dataset.

\subsection{CFD simulation setup}
\subsubsection{Geometry}
The simulated domain represents a mechanically ventilated classroom with dimensions of 19m × 13m × 3.5m, see Figure~\ref{fig:cadmodel} (a). 
The ventilation system consists of 18 rectangular inlet vents, each measuring 0.1349m in width and 0.3048m in height, and 2 ceiling-mounted outlet vent. Fresh air is supplied through the inlets at prescribed velocity and angle conditions, and exhausted through the outlets, forming a displacement ventilation flow pattern.

\subsubsection{Occupant modeling}
Occupants are collectively represented by a single, centrally positioned rectangular cuboid, visualized in Figure~\ref{fig:cadmodel}(a, right). This abstraction aggregates the influence of multiple seated individuals into a compact volume that approximates their combined occupied zone~\cite{chen2017modeling, d2023experimental}. 

CO$_2$ emissions from occupants are represented as a surface mass flux boundary condition and are modeled using the species transport equation~\cite{bulinska2014experimentally, mahmoud2024numerical}. 
To simulate varying occupancy levels (ranging from 10 to 80 people), we prescribe a total CO$_2$ mass flux of $n_p \times 0.00012$~kg/s on the cuboid surface, where $n_p$ is the number of occupants. This corresponds to a constant breathing rate of 6~L/min per person, following the simplified exhalation model in~\cite{bulinska2014experimentally}. A constant surface temperature is applied to the cuboid. 


\subsubsection{Mesh}
The fluid domain is discretized into approximately 0.245 million tetrahedral elements to resolve airflow and CO$_2$ transport dynamics. Local mesh refinement is applied near critical regions, including the occupant cuboid, inlet and outlet vents, and wall boundaries, to capture steep gradients in velocity and scalar fields. Mesh quality is assessed using standard metrics: the minimum orthogonal quality is 0.172, and the maximum aspect ratio is 43.33, indicating an acceptable mesh for transient indoor airflow simulations.

\subsubsection{Numerical models and boundary conditions}
Airflow and CO$_2$ dynamics within the classroom are simulated using the commercial CFD software ANSYS Fluent~\cite{manual2009ansys}. The incompressible Navier–Stokes equations are solved to capture airflow behavior, coupled with species transport equations to model the distribution of CO$_2$, O$_2$, H$_2$O, and N$_2$. Turbulence is modeled using the $k$–$\omega$ SST (Shear Stress Transport) model, which is well-suited for predicting indoor near-wall flow behavior~\cite{abuhegazy2020numerical}. The governing equations are discretized using the finite volume method, with second-order schemes for both momentum and species transport. The energy equation is also activated to solve for temperature.

The boundary surfaces include walls, inlet and outlet vents, and the occupant surface. A summary of boundary conditions is provided in Table~\ref{tab:boundary_conditions}. Inlets are modeled as velocity boundaries with prescribed temperature ($T_{\text{inlet}}$), CO$_2$ concentration ($C_{\text{inlet}}$), velocity magnitude ($m^r$), and velocity angle ($m^a$)~\cite{bulinska2014experimentally, mahmoud2024numerical}. 
To introduce airflow variability, the inlet velocities and angles for vent groups are randomly sampled from uniform distributions. A maximum velocity of 3.24~m/s corresponds to 10 air changes per hour (ACH), based on operational data. 
Outlets are modeled as pressure boundaries, and all walls are treated as no-slip, adiabatic surfaces with fixed temperature.


\begin{table}[htbp]
\centering
\caption{Boundary conditions.}
\color{donecolor}{
\begin{tabular}{lp{5cm}}
\toprule
\textbf{Boundary surface} & \textbf{Boundary conditions} \\
\midrule
Inlet vents & velocity intlet, $C_{\text{inlet}}=$400ppm, $T_{\text{inlet}}=$20$^\circ$C, $m^r \sim U[0.324, 3.24]$m/s, $m^a \sim U[45, 135]^\circ$\\
Outlet vents & Pressure outlet \\
Walls & No-slip and adiabatic walls, $T_{\text{wall}}=21^\circ C$ \\
Occupant surface & mass flow inlet, $C_{\text{occupant}}=40000$ppm,  $T_{\text{occupant}}=$25$^\circ$C~\cite{bulinska2014experimentally} \\
\bottomrule
\end{tabular}}
\label{tab:boundary_conditions}
\end{table}

\paragraph{Remark 2.} 
\seconddone{
The CFD simulations used in this work follow standard modeling practices for indoor airflow~\cite{d2023experimental,bulinska2014experimentally,weekly2015modeling}, and are designed to provide consistent and reproducible training data for future research. 
The simulation setup assumes constant air density (incompressible flow), omitting the buoyancy effects. 
This simplification is reasonable in mechanically ventilated classrooms where forced ventilation dominates airflow. Nevertheless, we acknowledge that in densely occupied spaces, buoyancy forces may contribute to airflow patterns.
Extending the CFD setup to incorporate buoyancy remains an important direction for future work, and our framework is readily adaptable to such enhanced simulations as well as to other CFD configurations and building control systems.
}

\color{black}
\subsection{Bear-CFD Dataset}
Bear-CFD dataset is generated through a structured CFD simulation workflow designed to capture the spatiotemporal dynamics of indoor CO$_2$ levels under varying ventilation and occupancy conditions. The dataset is publicly available at \url{https://ucsdsmartbuilding.github.io/CFD-DATA.html}.

The simulations include both steady-state and transient flow cases. \done{Specifically, we run 10 steady-state simulations under different boundary conditions. These steady-state results are then used as \textit{initial conditions} for 300 transient simulations, each lasting 30 minutes of physical time and resolved at 10-second time steps. 
This initialization strategy ensures that each transient simulation begins from a realistic state shaped by prior control conditions. Without this step, starting from uniform or arbitrary states (e.g., same CO$_2$ everywhere) could introduce artificial transients unrelated to the actual control inputs.} 
Simulation outputs are saved every 30 seconds, resulting in 60 time steps per transient case. Transient simulations capture the temporal evolution of airflow dynamics, reflecting unsteady effects influenced by varying boundary conditions. We leverage this transient data to train our neural operators. 

To generate the dataset, for each simulation, key control parameters, including supply airflow rates ($m_i^r$) and airflow angles ($m_i^a$) for the $i$-th group of inlet vents ($i = 1,\ldots,6$), as well as the occupant count ($n_p$)—are independently sampled from uniform distributions:
\begin{equation}\label{eq:uniform_control}
\begin{aligned}
   & m_i^r \sim U[\underline{m^r}, \overline{m^r}], \, m_{i}^a \sim U[45^\circ, 135^\circ], \,\,i \in [1,\ldots, 6], \\
   & n_p \sim U[10, 80], 
\end{aligned}
\end{equation}
where $m_i^r$ is the airflow rate for the $i$-th group of vents, bounded between $\underline{m^r}$ = 0.324m/s (10\% of maximum) and $\overline{m^r}$=3.24m/s; $m_i^a$ is the airflow angle of the $i$-th group of vents, spanning $45^\circ$ to $135^\circ$; and $n_p$ is the number of occupants in the classroom.
\done{Each transient simulation follows this two-phase procedure: (1) Steady-state initialization: A random initial condition is selected from the 10 steady-state simulations. (2) Transient simulation: Control parameters ($m_i^r$, $m_i^a$) and occupancy $n_p$ are sampled. The simulation runs for 30 minutes, with CO$_2$ and airflow fields recorded at 30-second intervals over $\overline{T} = 60$ time steps.
} The CO$_2$ concentrations were monitored at two critical planes:
\begin{itemize}
    \item HVAC surface: A horizontal plane at 2.9-meter height near the ventilation inlets, capturing the supply and returning air quality.
    \item People surface: A horizontal plane at 1.6-meter height (average breathing height for standing adults), representing air quality at occupant exposure levels~\cite{ashrae1992standard55}.
\end{itemize}
In our CFD simulation, the sitting surface is near the occupancy boundary, leading to potential inaccuracies from numerical artifacts and boundary effects. Thus, we focused on heights where airflow and CO$_2$ dispersion are more reliably captured. 

The dataset is distributed in both ANSYS FLUENT's native format (.cas and .dat files) and Python pickle (.pkl) format. The FLUENT files preserve the complete simulation environment and solution data, while the pickle format enables efficient programmatic access to the numerical results through standard Python libraries. The dataset includes spatiotemporal fields of CO$_2$ concentrations, airflow rates, vent angles, and occupancy, as detailed in Table~\ref{tab:data_fields}.

\begin{table}[htbp]
\centering
\caption{Data fields in the BEAR-CFD dataset pickle (.pkl) format.}
\label{tab:data_fields}
\begin{tabular}{p{2.cm}p{5cm}}
\hline
Field & Description \\ \hline 
HVAC surface & ndarray ($N_{\text{hvac}}$, 3), spatial coordinates of grid points on HVAC surfaces \\ 
CO$_2$-HVAC & ndarray ($N_{\text{hvac}}$,$\overline{T}$), CO$_2$ concentration time series at HVAC surface\\ 
People surface & ndarray ($N_{\text{people}}$, 3), spatial coordinates of grid points on people  surfaces \\ 
CO$_2$-People & ndarray ($N_{\text{people}}$,$\overline{T}$), CO$_2$ concentration time series at people surface\\ 
steady case & int, Identifier for the initial steady-state condition used in the simulation\\
$n_p$ & int, number of occupant \\
$m_i^r$ & float, airflow rate (m/s) for $i$-th group of vents \\
$m_i^a$ & float, angle ($^\circ$) for $i$-th group of vents \\ \hline
\end{tabular}
\end{table}

%% file: method.tex
\begin{figure*}[t]
    \centering
    \includegraphics[width=0.95\linewidth]{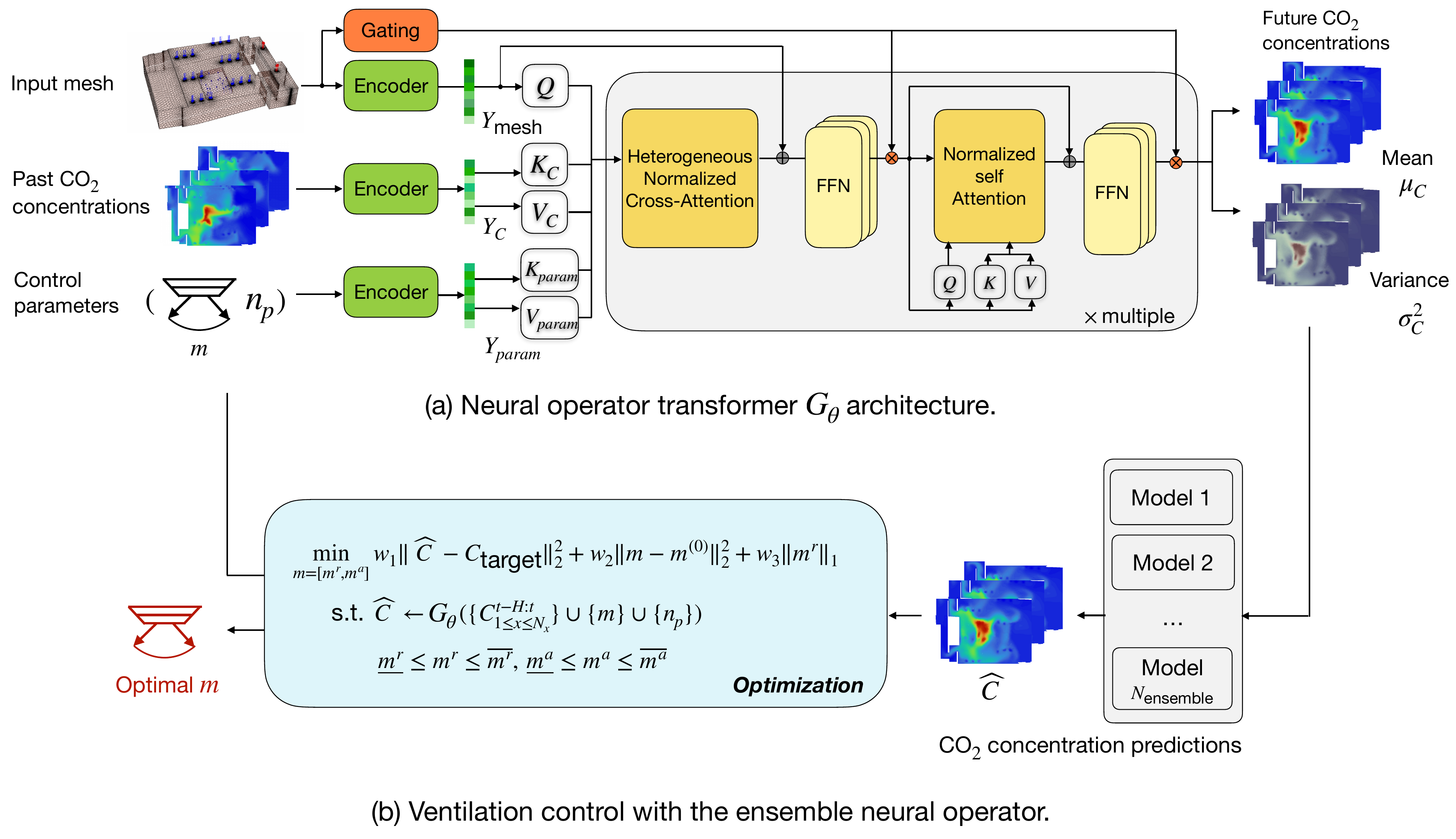}
    \caption{Overview of the proposed data-driven operator learning framework for energy-efficient ventilation control. The framework consists of two phases: (1) the learning phase, where neural operator transformers are trained to map past air field data and ventilation control parameters to future air field evolution, and (2) the control phase, where the trained ensemble neural operator is integrated into an optimization framework to solve the ventilation control problem. This approach enables real-time optimization of airflow supply rates and vent angles while maintaining air quality standards.}
    \label{fig:framework}
\end{figure*}

In this work, we propose a data-driven operator learning framework to model the indoor air quality and optimize ventilation control for energy efficiency. 
The core component is an ensemble neural operator transformer architecture, $\mathcal{G}_\theta$, shown in Figure~\ref{fig:framework}. In figure~\ref{fig:framework}(a), the neural operator architecture enables fast building CFD simulations by processing multiple inputs: query points, supply airflow rates,  airflow angles, and historical CO$_2$ concentrations. The trained operator learning model uses these inputs to predict future CO$_2$, which are then utilized to determine optimal ventilation control parameters by solving the optimization problem defined in Equation~\eqref{eq:problem}.
In the following subsections, we will first explain the concept of operator neural transformer, then detail describe the control algorithm. 

\subsection{Ensemble neural operator transformer}
Existing operator learning approaches, while effective in many applications, often struggle with limited training data. To address this limitation, we enhance the General Neural Operator Transformer (GNOT)~\cite{hao2023gnot} by ensemble learning~\cite{lakshminarayanan2017simple}. The network architecture of the proposed model together with the control is illustrated in Figure~\ref{fig:framework}. In the following subsections, we will describe the input encoding and ensemble learning for GNOT.

\subsubsection{Input encoding}
The model takes the input mesh, historical CO$_2$ concentrations, and control parameters (including ventilation control actions and the number of occupants), as inputs. To accommodate these heterogeneous inputs, a general encoder, highlighted in green in Figure~\ref{fig:framework}, is employed to transform them into the feature embedding $Y \in \mathbb{R}^{N \times n_e}$, where $N$ denotes an arbitrary number of input elements and $n_e$ is the embedding dimension. The model employs simple multilayer perceptrons (MLPs), denoted as $f_{w1}, f_{w2}, f_{w3}$, to map each type of input to its corresponding embedding. 

\begin{itemize}
    \item Input mesh: 
    A MLP maps the mesh points to query embeddings $Y_{\text{mesh}} = (f_{w1}(x_i))_{1 \leq i \leq N_x} \in \R^{N_x \times n_e}$.
    \item Past CO$_2$ concentrations: At time $t$, we have $\{x_i, c_i\}_{1 \leq i \leq N_x}$, where $c_i = C_i^{t-H:t}$ represents the historical CO$_2$ levels at position $x_i$. A MLP encodes both positions and concentrations to produce the feature $Y_C = (f_{w2}(x_i, c_i))_{1 \leq i \leq N_x} \in \mathbb{R}^{N_x \times n_e}$.
    \item Control parameters: A MLP encodes control parameters $[m, n_p] \in \R^{13}$ into embeddings $Y_{\text{param}} = f_{w3}([m, n_p]) \in \mathbb{R}^{1 \cdot n_e}$.
\end{itemize}

\subsubsection{Ensemble learning of GNOT}
As shown in Figure \ref{fig:framework}(a), GNOT begins by encoding input features and updating them using a heterogeneous normalized cross-attention layer, followed by a normalized self-attention layer to refine representations. To effectively capture spatial heterogeneity, GNOT incorporates a geometric gating mechanism that leverages the query point coordinates to compute a weighted combination of expert feed-forward networks (FFNs). The model stacks $N$ such attention blocks to produce the final predictions. 

While GNOT is originally designed to output only the mean of the prediction, we enhance its robustness by introducing an ensemble-based extension. 
Instead of training GNOT to minimize mean squared error or relative error, we modify it to predict a probability distribution for future CO$_2$  concentrations, characterized by the mean ($\mu_C$) and variance ($\sigma^2_C$). Specifically, the model predicts:
\begin{subequations}\label{eq:meanvariance}
\begin{align}
& \mu_C, \sigma^2_C = \mathcal{G}_\theta (\mathcal{A}), \\
& \mu_C   = \mathbb{E}[\widehat{C}_{1\leq i \leq N_x}^{t+1:t+T}], \,\,
\sigma^2_C   = \text{Var}[\widehat{C}_{1\leq i \leq N_x}^{t+1:t+T}],
\end{align}
\end{subequations}
where $[\mu_C]_i^t, [\sigma^2_C]_i^t$ represents the mean and variance at location $i$ and time step $t$.
The model is trained using the Negative Log-Likelihood (NLL) loss, defined as:
\begin{equation}\label{eq:nll_loss}
\mathcal{L} = \frac{1}{N_x T} \sum_{i=1}^{N_x} \sum_{k=1}^T \left(\frac{\log(2\pi[\sigma^2_C]_i^{t+k})}{2}  + \frac{(C_i^{t+k} - [\mu_C]_i^{t+k})^2}{2[\sigma^2_C]_i^{t+k}} \right)
\end{equation}
where $C_i^{t+k} \in \R$ is the true value at $x_i$ and time $t+k$. By minimizing the NLL loss, the model learns to jointly optimize the mean and variance predictions, effectively mitigating overfitting~\cite{lakshminarayanan2017simple}. 

We train an ensemble of neural operator transformers, with the final prediction obtained by averaging outputs from multiple independently trained models. Let $\mu^{(n)}_C$ denote the mean prediction of the $n$-th model. The ensemble prediction is then computed as,
\begin{subequations}\label{eq:ensemble_pred}
\begin{align}
    & \mu_C^{\text{ensemble}}  = \frac{1}{N_{\text{ensemble}}}
    \sum_{n=1}^{N_{\text{ensemble}}} \mu_C^{(n)}, \\
    & \mu_C^{(n)}, (\sigma^2)_C^{(n)} = \mathcal{G}_{\theta_n} (\mathcal{A}), 
\end{align}
\end{subequations}
where $N_{\text{ensemble}}$ represents the number of models in the ensemble, $\theta_n$ is the model parameters for $n$-th trained neural operator. This ensemble approach not only improves prediction accuracy but also enhances the reliability of ventilation control, as demonstrated in our experiments.

\subsection{Control algorithm}
With the learned ensemble neural operator transformer model, we are ready to solve the building control problem in \eqref{eq:problem}. Recall that the objective function of the building control problem is defined as:
\begin{equation}\label{eq:control_loss}
    \mathcal{L}(m) = w_1 \|\widehat{C}-C_{\text{target}}\|_2^2 + w_2 \|m-m^{(0)}\|_2^2 + w_3 T \|m^r\|_1
\end{equation}
where the first term $\|\widehat{C}-C_{\text{target}}\|_2^2$ is defined as,
$$\|\widehat{C}-C_{\text{target}}\|_2^2 = \frac{1}{N_x T}\sum_{k=1}^T \sum_{i=1}^{N_x} (\widehat{C}_i^{t+k} - {C_{\text{target}}}_i^{t+k})^2$$
and $w_1, w_2, w_3$ are weighting coefficients. $\widehat{C}$ represents the predicted CO$_2$ concentrations generated by the ensemble neural operator~\eqref{eq:ensemble_pred}:
\begin{equation}\label{eq:ensemble_pred2}
    \widehat{C} = \mu_C^{\text{ensemble}}.
\end{equation}

To solve the building control problem in \eqref{eq:problem}, we leverage the differentiability of the neural operator $\mathcal{G}_\theta$ and propose a gradient-based method to update the control actions for ventilation. To ensure that the control vector $m$ remains within feasible bounds~\eqref{eq:problem_c2}, we use a projected gradient descent method. Specifically, after computing the gradient of the objective function with respect to $m$, the control vector is updated and then clipped to satisfy the predefined bounds. Our optimization procedure is summarized in Algorithm~\ref{alg}. 

\begin{algorithm}[t]
\caption{Algorithm for solving~\eqref{eq:problem}}\label{alg}
\begin{algorithmic}[1]
\Require  Neural operator transformers $\mathcal{G}_\theta$, \\ 
$\quad \quad \,\, $ Control inputs $C^{t-H:t}_{1 \leq i \leq N_x}, n_p$
\Ensure $m^{(0)}$ \Comment{initial control actions}
\For{$ite = 0, 1, \ldots \text{MaxIte}$}
\State \textbf{obtain} future CO$_2$ predictions $\widehat{C}$~\eqref{eq:ensemble_pred}\eqref{eq:ensemble_pred2}
\State \textbf{evaluate} loss $\mathcal{L}(m)$~\eqref{eq:control_loss}
\State \textbf{compute} the gradient $\nabla \mathcal{L}(m)$
\State \textbf{update} the control vector with step size $\eta$:
$$m^{(ite+1)} \leftarrow m^{(ite)} - \eta \nabla \mathcal{L}(m)$$
\State \textbf{project} the update $m$ to satisfy the box constraints~\eqref{eq:problem_c2}
\EndFor
\State \textbf{Return $m$}
\end{algorithmic}
\end{algorithm}

%% file: experiment.tex
In this section, we evaluate the performance of the proposed framework through two experiments: (1) learning experiments to assess the accuracy of the ensemble neural operator, and (2) control experiments to evaluate the effectiveness of our data-driven ventilation control framework. The source code, input
data, and trained models from all experiments are available on GitHub\footnote{\url{https://github.com/alwaysbyx/BuildingControlCFD}}. All experiments are conducted on NVIDIA GeForce RTX 2080 Ti GPUs.

\subsection{Learning results}\label{sec:learning_result}
We train our ensemble neural operator transformer to predict CO$_2$ levels on the people surface. The number of query points is $N_x=7462$, and we select $F=12$ and $T=6$, meaning that the model utilizes data from the past 12 time steps (equivalent to 6 minutes) to forecast CO$_2$ levels over the next 6 time steps (3 minutes). The dataset is divided into an 80\% training set and a 20\% testing set to evaluate model performance. We train $N_{\text{ensemble}}=5$ independent models and compute the mean of their predictions as the final output. We use the AdamW optimizer with a cyclical learning rate schedule, and each model is trained for 200 epochs.  
\begin{figure*}
    \centering
    \includegraphics[width=0.9\linewidth]{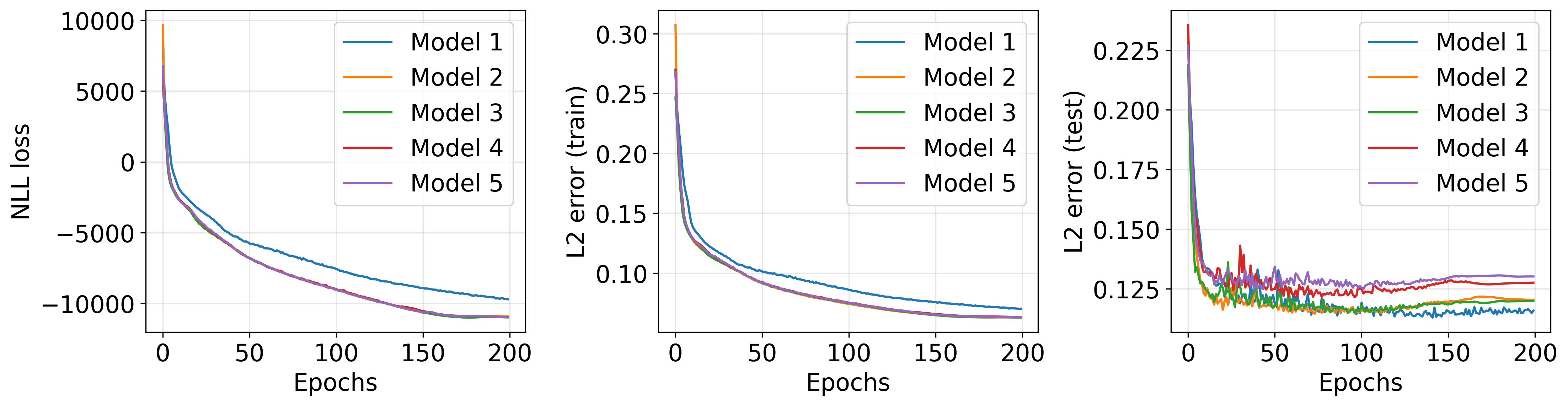}
    \caption{Training loss (NLL loss) during training (left) and the $l_2$ error for the training (middle) and test sets (right). The $l_2$ error~\eqref{eq:relative_error} is computed based on the ground truth and the model's mean prediction output.}
    \label{fig:training_curve}
\end{figure*}
\begin{table*}
\begin{tabular}{lllllll} \hline
         & Model 1 & Model 2 & Model 3 & Model 4 & Model 5 & Ensemble \\ \hline 
$l_2$ error (Train) &  6.35\%    &   6.33\%      &   6.33\%      &    6.35\%     &    6.33\%     &     \textbf{5.9\%}     \\
$l_2$ error (Test)  &   12.09\%  &   11.83\%      &   11.82\%      &    12.74\%     &   13.01\%      &   \textbf{10.90\%}     \\ \hline
\end{tabular}
\caption{The $l_2$ error for five independently trained neural operator transformer models (Model 1 to Model 5) and their ensemble.}
\label{tab:learn}
\end{table*}
\begin{figure*}[thb]
    \centering
    \includegraphics[width=\linewidth]{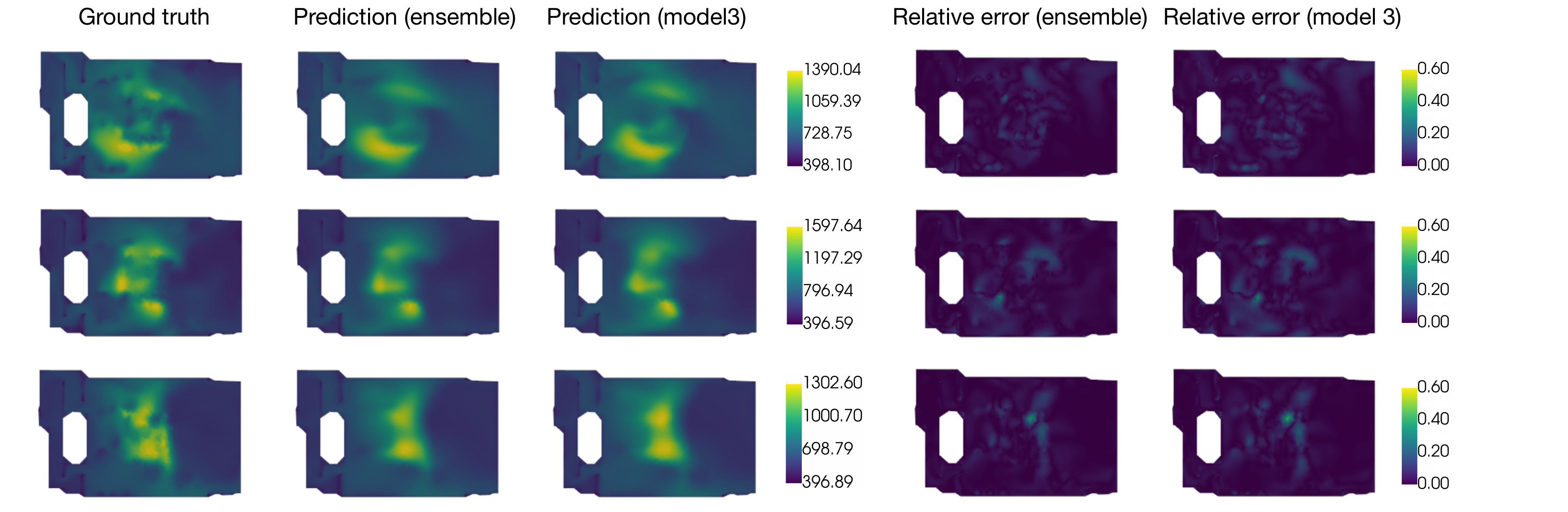}
    \caption{Operator Learning: Visualization of the ground truth, corresponding predictions from the ensemble model and Model 3, and the relative errors between the ground truth and predictions at the final time step. }
    \label{fig:predict}
\end{figure*}

To evaluate model performance, we use the average $l_2$ relative error as the primary metric. Let $C^{(d)}, \widehat{C^{(d)}} \in \R^{N_x \times T}$  denote the ground truth and the predicted \textit{mean} future CO$_2$  concentrations for the $d$-th sample, respectively, and let $D$ represent the total dataset size. The error is defined as:
\begin{equation}\label{eq:relative_error}
l_2 = \frac{1}{D} \sum_{d=1}^{D} \frac{\|\widehat{C^{(d)}} - C^{(d)}\|_2}{\|C^{(d)}\|_2}. 
\end{equation}
Figure~\ref{fig:training_curve} shows the training convergence and test performance of the five neural operator transformers over 200 epochs. The left plot presents the NLL loss~\eqref{eq:nll_loss}, while the right shows the 
$l_2$ relative error on both training and test sets.
Table~\ref{tab:learn} list the $l_2$ error metrics for the five trained models (Model 1 to Model 5) and the ensemble model, evaluated on both the training and testing datasets. Notably, the ensemble model achieves the lowest errors, with a training $l_2$ error of 5.9\% and a testing $l_2$ error of 10.90\%. The ensemble model consistently outperforms the individual models, leading to more robust and accurate predictions. The performance improvement can be attributed to the ensemble model's ability to mitigate potential overfitting of individual models by averaging out their prediction errors. 

Figure~\ref{fig:predict} shows the ground truth CO$_2$ concentration fields with predictions from the ensemble model and Model 3 (best-performing individual model), along with their relative errors for three test cases. Overall, the neural operator framework shows strong capability in predicting complex spatial CO$_2$ distributions under varying control parameters.
The ensemble model achieves lower relative errors and more accurately captures spatial patterns compared to Model 3. In Section~\ref{sec:ensemble}, we further demonstrate its effectiveness in activating control actions and optimizing the air quality.

\begin{table*}[t]
\centering
\begin{tabular}{llccccccc}
\hline
Case & Control Strategy & \multicolumn{2}{c}{Mean CO$_2$ (ppm)} & \multicolumn{2}{c}{Peak CO$_2$ (ppm)} & \multicolumn{2}{c}{CO$_2$ > 1200ppm (\%)} & Energy \\
& & Average & Final Step & Average & Final Step & Average & Final Step & Consumption (\%)\\ 
\hline
\multirow{4}{*}{1} 
& Max Control          & 565.6 & 534.0 & 895.7 & 854.2 & 0.00 & 0.00 & 100.0 \\
& Baseline Control     & 616.1 & 612.3 & 1203.0 & 1143.6 & 1.32 & 0.00 & 50.0 \\
& Rule-based Control   & 567.4 & 533.5 & 898.1 & 807.9 & 0.00 & 0.00 & 93.8 \\
& Data-driven(DL-Avg)    & 604.4 & 595.2 & 1021.6 & 971.6 & 0.00 & 0.00 & 83.3 \\
& Data-driven(DL-ROM)  & 626.2 & 635.8 & 1300.1 & 1326.7 & 8.34 & 10.56 & 44.6 \\
& Ours                 & 600.7 & 587.9 & 1060.2 & 1004.0 & \textbf{0.00} & \textbf{0.00} & \textbf{65.8} \\ 
\hline
\multirow{4}{*}{2} 
& Max Control          & 546.7 & 512.2 & 800.0 & 741.3 & 0.00 & 0.00 & 100.0 \\
& Baseline Control     & 599.9 & 595.9 & 1097.3 & 1108.4 & 0.00 & 0.00 & 50.0 \\
& Rule-based Control   & 593.7 & 584.6 & 1057.6 & 1059.3 & 0.00 & 0.00 & 56.2 \\
& Data-driven(DL-Avg)     & 584.7 & 562.1 & 912.7 & 914.6 & 0.00 & 0.00 & 83.0 \\ 
& Data-driven(DL-ROM)  & 602.7 & 600.1 & 1062.2 & 1066.2 & 0.00 & 0.00 & 50.8 \\
& Ours                 & 605.0 & 600.0 & 1044.1 & 881.4 & \textbf{0.00} & \textbf{0.00} & \textbf{43.8} \\
\hline
\multirow{4}{*}{3} 
& Max Control          & 532.3 & 497.6 & 754.0 & 692.0 & 0.00 & 0.00 & 100.0 \\
& Baseline Control     & 585.2 & 578.0 & 1034.2 & 1013.9 & \textbf{0.00} & \textbf{0.00} & \textbf{50.0} \\
& Rule-based Control   & 650.7 & 712.7 & 1361.4 & 1623.1 & 15.67 & 35.26 & 18.8 \\
& Data-driven(DL-Avg)    & 576.9 & 566.0 & 946.1 & 978.3 & 0.00 & 0.00 & 66.1 \\
& Data-driven(DL-ROM)  & 590.3 & 590.2 & 1121.6 & 1208.4 & 0.13 & 0.70 & 49.4 \\
& Ours                 & 589.4 & 585.4 & 984.4 & 949.8 & 0.00 & 0.00 & 55.2 \\ 
\hline
\end{tabular}
\caption{Comparison of CO$_2$ metrics (mean, peak, and violation percentages) and energy consumption across different control strategies. "Average" refers to the temporal mean over the simulation period, while "Final Step" refers to the CO$_2$ level at the end of the simulation. Our approach achieves significant ventilation energy savings compared to the other control approaches while maintaining acceptable CO$_2$ violation levels. 
}
\label{tab:control}
\end{table*}

\subsection{Ventilation control results}\label{sec:venti1}
We now evaluate the control performance under three occupancy scenarios: high ($n_p=75$), medium ($n_p=45$), and low ($n_p=15$). 
According to ASHRAE Standard 62.1~\cite{ashrae20102standard621}, classrooms typically require 4–6 air changes per hour (ACH) to ensure adequate ventilation and indoor air quality. For all scenarios in this study, the initial control actions are initialized at 5 ACH (within the recommended range) for supply vents, with inlet angles fixed at a 90$^\circ$ downward orientation to align with conventional HVAC configurations.
We evaluate the following control strategies:
\begin{itemize}
\item \textbf{Max Control}: The airflow rate is set to its maximum value, with the inlet angle fixed at 90$^\circ$ downward. 
\item \textbf{Baseline Control}: The control actions determined based on the ASHRAE standard as 5 ACH with 90$^\circ$ downward angle, serving as a baseline for comparison.
\item \textbf{Rule-based Control}: The airflow rate is set proportionally to the occupancy level, calculated as the number of occupants divided by the maximum occupancy, with 90$^\circ$ downward angle.
\item \textbf{\done{Data-driven (DL-Avg)}}: We employ a neural network to predict the average CO$_2$ concentration at the center of the occupancy surface. Ventilation control actions are optimized based on \eqref{eq:problem} with the prediction.
\item \textbf{Data-driven (DL-ROM)}: \done{We implement a state-of-the-art Deep Learning–based Reduced-Order Modeling approach(DL-ROM)~\cite{pant2021deep} to approximate the spatiotemporal CO$_2$ distribution. The resulting surrogate model is integrated into our ventilation control optimization framework \eqref{eq:problem} to optimize control actions. }
\item \textbf{Ours}: \done{We use the proposed ensemble neural operator transformer to model the spatiotemporal dynamics of CO$_2$ concentrations. The learned neural operator is then used within the same ventilation control optimization framework in \eqref{eq:problem} to optimize control actions.}
\end{itemize}

\done{For our optimization solver, we choose $C_{\text{target}}=400$ (ppm) to ensure that the CO$_2$ concentration loss is more broadly activated across the spatial domain, enabling more effective optimization with respective to the control actions.
We choose $w_1=1, w_2=0.4$ and $w_3=0.12$ through empirical tuning to balance air quality, energy consumption, and control smoothness. These weights are used consistently across all control experiments. }

We validate the three control strategies using CFD simulations. Performance is measured using three metrics: mean CO$_2$ (ppm), peak CO$_2$ (ppm), and CO$_2$ violation (\%), defined as $\frac{\text{Peak CO}_2 - 1200}{1200}$.
As suggested by~\cite{zhang2023model}, 1200 ppm is the maximum acceptable CO$_2$ level for human health. For each metric, we record both the temporal average over the simulation period and the value at the final time step.
Energy consumption is modeled as the percentage of the maximum power required to operate the ventilation system at its maximum airflow rate, providing a \emph{normalized} measure of energy usage relative to the system's peak capacity:
\begin{equation}\label{eq:energy }
E(m) = \frac{1}{6} \sum_{i=1}^6 m_i^r / \overline{m^r} \times 100\%,
\end{equation}
where $m_i^r$ represents the ventilation rate for the $i$-th vent, and $\overline{m^r}=3.24$ (m/s) is the maximum ventilation rate.

                  
                   
                

The results are summarized in Table~\ref{tab:control}. The experimental results demonstrate the effectiveness of the proposed control strategy (Ours) in balancing air quality and energy consumption in the first two cases. In Case~3, under the low-occupancy scenario, the optimization favors maintaining higher air quality, leading to slightly higher energy consumption than the baseline control.
Compared to the Max control strategy, our method achieves comparable air quality performance while reducing energy usage by 34–56\%. Compared to the baseline control, our approach effectively lowers CO$_2$ levels and adheres to air quality constraints with slight energy increase. 
Rule-based control suffers from CO$_2$ violations (e.g., 35.2\% final-step violation in Case 3) due to its reliance on well-chosen static operation schedules. Unlike rule-based systems, which require manual fine-tuning of thresholds for different room layouts and occupancy levels, our method autonomously adapts to operation conditions and reduces energy consumption by 12-28\% in Cases 1-2 while maintaining safe CO$_2$ levels. 

In addition, our method outperforms data-driven control (DL-Avg) and (DL-ROM) by maintaining good energy efficiency without incurring any violations of the air quality constraints. 
While the DL-Avg approach can optimize ventilation locally, it has two key limitations: (1) it primarily predict average CO$_2$ concentrations rather than spatial variations, and (2) it cannot effectively map control actions to real-time CO$_2$ distributions across different zones. 
\done{While the DL-ROM approach captures spatial-temporal airflow dynamics, it is less effective at modeling the relationship between control inputs and the resulting air distributions compared to our method. For completeness, we present the learning results of DL-ROM in Appendix~\ref{ap:DL-ROM}.}

\subsection{Multi-step ventilation control results}
To assess the proposed operator learning model in realistic control scenarios, we extend the single-step optimization in Section~\ref{sec:venti1} to a multi-step planning horizon.
Specifically, a 30-minute ventilation horizon is considered with 10 planning steps, each optimizing control over the next 6 time steps (3 minutes), yielding 60 control steps in total.
The initial step ($t=0$) is initialized using historical CO$_2$ data from Scenario 1 in Subsection~\ref{sec:venti1}, while subsequent steps use the predicted CO$_2$ levels by recursively rolling out the learned operator model with previously optimized controls.

\begin{figure}
    \centering
    \includegraphics[width=1.0\linewidth]{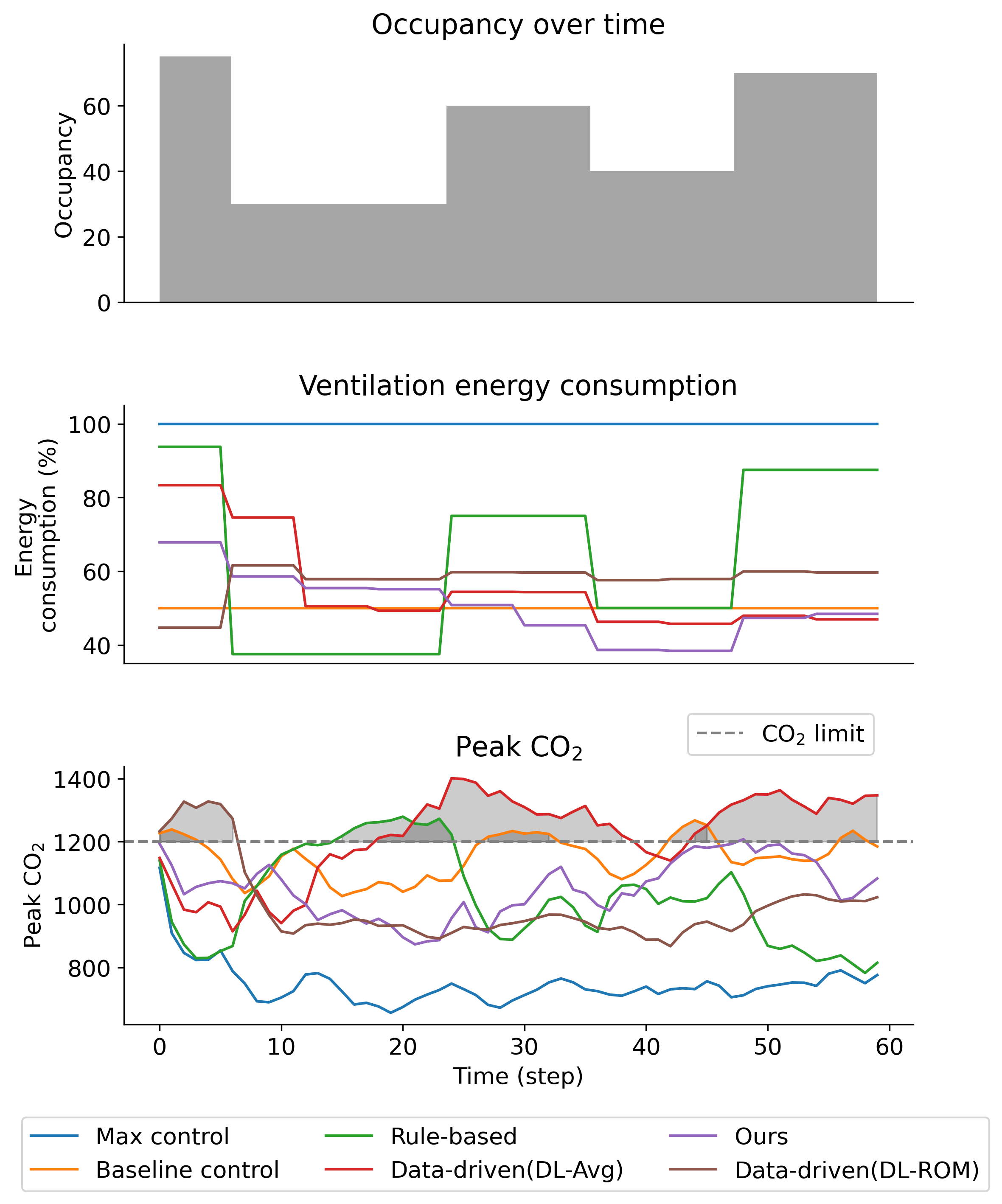}
    \caption{Comparison of occupancy profile, energy consumption, and peak CO$_2$ concentration over a 60-step period. The top panel illustrates dynamic occupancy changes, the middle panel shows corresponding energy consumption as a percentage of the maximum, and the bottom panel depicts CO$_2$ concentration levels with the threshold limit (1200 ppm) indicated by a dashed line. \vspace{-12pt}}
    \label{fig:mpc}
\end{figure}

\begin{figure}[b]
    \centering
    \includegraphics[width=0.9\linewidth]{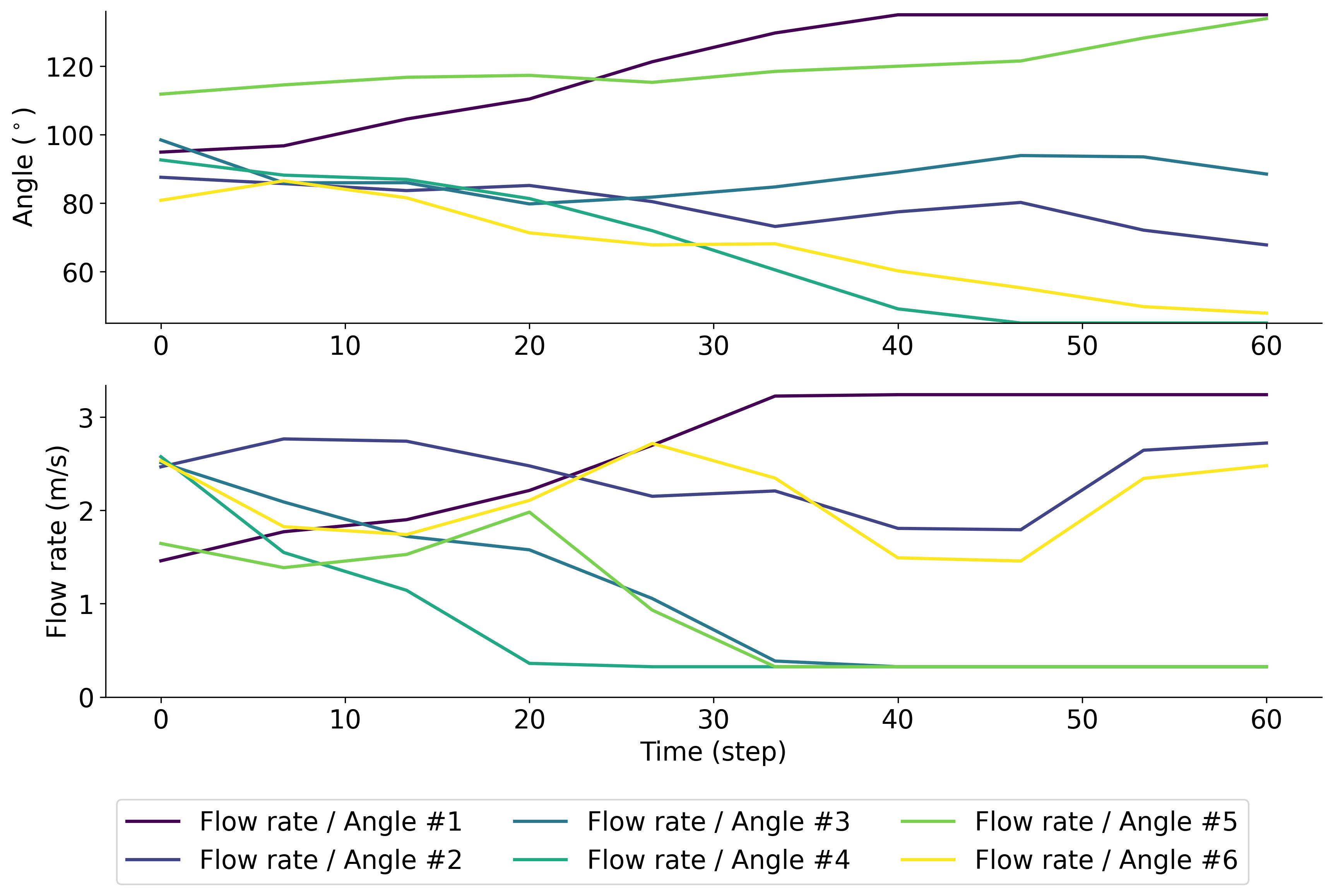}
    \caption{\done{Time series of control actions of our approach. Top plot shows the airflow angles and bottom shows airflow rates for 6 group of inlet vents under dynamic occupancy levels. }}
    \label{fig:action}
\end{figure}

Figure~\ref{fig:mpc} illustrates the performance of the proposed control strategy in comparison to other control methods. The occupancy profile (top) shows significant variations over time. 
Our strategy dynamically adjusts ventilation in response to occupancy levels. At the beginning, when occupancy is high, the controller increases airflow to mitigate CO$_2$ accumulation. Although occupancy drops shortly after, the controller maintains a moderately high ventilation rate to clear the residual CO$_2$ from the earlier crowded period. In contrast, during later low-occupancy intervals, lower accumulated CO$_2$ allows the system to reduce airflow further, resulting in greater energy savings.

\done{Over the control period, total energy consumption for Max, Baseline, Rule-based, Data-driven (DL-Avg), Data-driven (DL-ROM), and our method is 100\%, 50\%, 63.1\%, 55.3\%, 57.6\%, and 50.6\%, respectively. In terms of CO$_2$ violations, Max control yields zero violations; Baseline, Rule-based, and DL-ROM incur 17, 10, and 7 violations; DL-Avg results in 37 violations and ours incurs only 1 violation.} The relatively poor performance of DL-Avg stems from its reliance on predicting only the average CO$_2$ concentration, without capturing the dynamic spatial shifts of peak concentrations over time. As a result, the control decisions fail to address localized high-risk areas, leading to both higher energy consumption and more CO$_2$ violations. 

\done{In addition, we plot the time series of control actions for the six groups of vents, under our approach in Figure~\ref{fig:action}. 
Notably, the airflow rates for vent groups 2 and 6 exhibit a strong correlation with occupancy, increasing during higher occupancy periods to enhance ventilation. Additionally, the controller consistently assigns the highest airflow to vent group 1, while groups 3, 4, and 5 maintain relatively low flow rates. This likely reflects spatial differences in ventilation effectiveness, with the optimization learns to prioritize vents that more effectively contribute to CO$_2$ removal.}

\subsection{Effectiveness of the ensemble model}
\label{sec:ensemble}
\begin{figure}[b]
    \centering
    \includegraphics[width=1.0\linewidth]{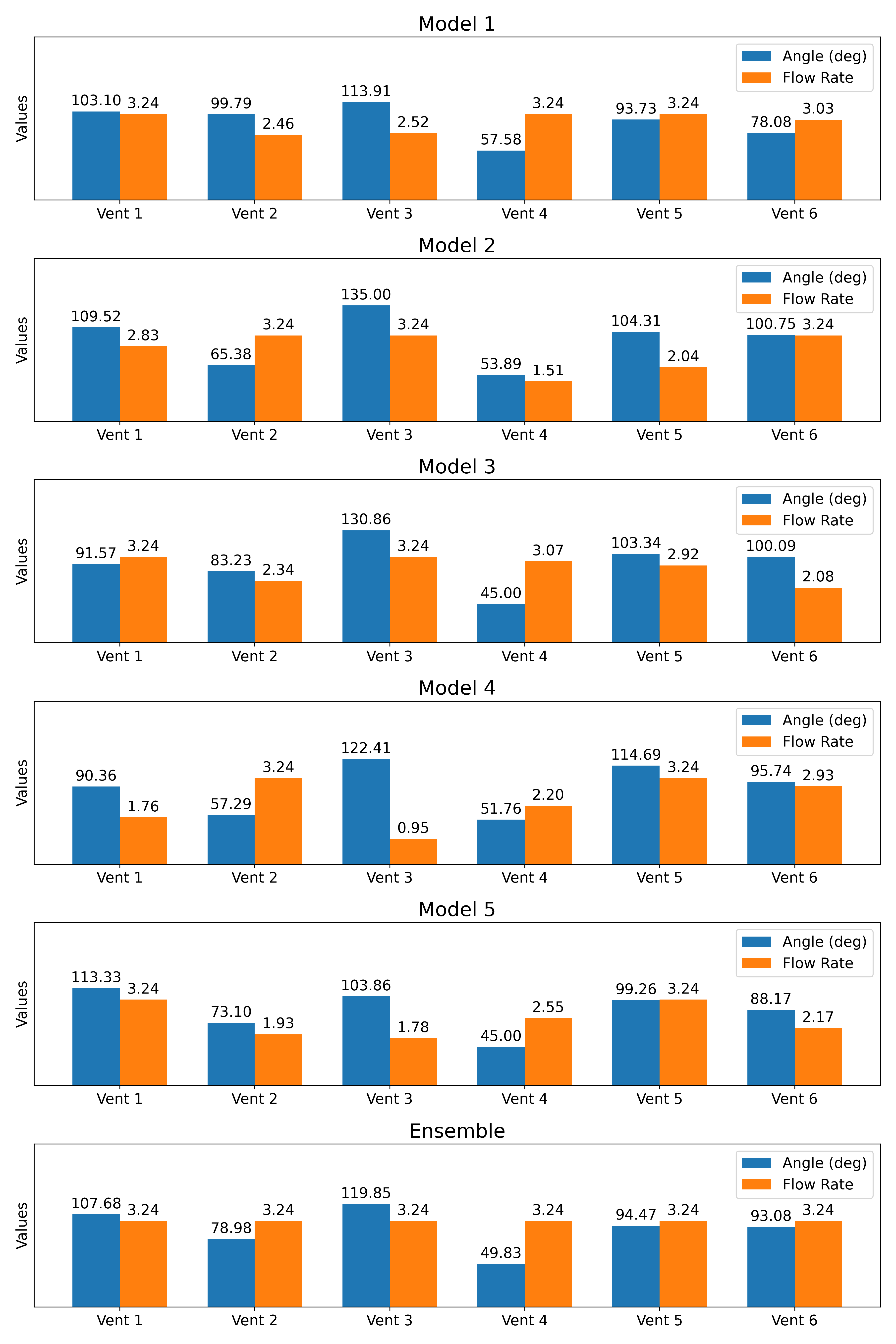}
    \caption{Control actions (flow rates and angles) for individual models (Model 1 to Model 5) and the ensemble model across six inlet vents for one case. The ensemble model consistently achieves maximum airflow rates for all inlet vents.}
    \label{fig:ensemble_control}
\end{figure}
In this subsection, we demonstrate the effectiveness of the ensemble model in control compared to individual models, serving as an ablation study. As shown in Section~\ref{sec:learning_result}, the ensemble model achieves the lowest prediction error among all models. Building on these results, we further illustrate that the ensemble model, which aggregates predictions from multiple independently trained neural operators, consistently outperforms individual models in the control stage. 

To gain more insights from the optimization results, we set the coefficient for energy consumption $w_3=0$ in the building control problem~\eqref{eq:problem}. This allows the control strategy to prioritize air quality without considering energy cost. 
We optimize the control actions for a case in which the control parameters are randomly selected, using both the individual models and the ensemble model.
The resulting control actions are illustrated in Figure~\ref{fig:ensemble_control}. We observe that the ensemble model consistently reaches maximum airflow rates for all inlet vents, which align with the optimal action in this test case as the energy cost coefficient $w_3 =0$. In contrast, the individual models exhibit variability in their control actions, with some models failing to achieve the maximum airflow rates across all vents.

\color{donecolor}

\subsection{Runtime analysis}
\label{sec:run_time}
To evaluate the computational efficiency of our proposed operator learning for building ventilation control, we analyze the runtime associated with each stage of the algorithmic pipeline, including the cost of generating CFD data, training time of the neural operator models, and performing inference during control deployment.

Table~\ref{tab:runtime_breakdown} summarizes the runtime of each component. The CFD simulations are conducted using ANSYS Fluent in parallel 6 cores, while neural operator training was performed using 2 RTX 2080 Ti GPUs. 
We observe that CFD simulation takes 1,253.7 seconds to compute the transient flow over six time steps (3 minutes), whereas the neural operator transformer completes the same task in just 0.005 seconds. This represents a remarkable speed-up of approximately {250,000} times compared to the CFD simulation. 
Although the learning phase requires moderate computational resources (the training of the five transformer models takes 16 GPU hours in total), the resulting neural operator enables rapid inference and real-time building control with the high-fidelity indoor fluid dynamics models.


\begin{table}[htbp]
\centering
\caption{Runtime of the operator learning pipeline.}
\label{tab:runtime_breakdown}
{\color{donecolor}
\begin{tabular}{p{4cm} p{3.2cm} }
\toprule
\textbf{Stage} & \textbf{Time Cost}  \\
\midrule
\textbf{Learning} & \\
CFD data generation & 3.5 CPU hours per simulation, totaling ~1100 CPU hours  \\
Neural operator training & 16 GPU hours \\
\midrule 
\textbf{Control} & \\
CFD simulation (Sec 5.2) & 1253.7 seconds  \\
CFD simulation (Sec 5.3) & 3.5 hours  \\
Inference (ours) & \textless 0.005 seconds \\
\bottomrule
\end{tabular}
}
\end{table}

\color{black}

%% file: appendix.tex
\section{Airflow dynamics modeling}
\label{ap:velocity}

Indoor air is modeled as an incompressible fluid consisting of four species, namely: oxygen (O$_2$), carbon dioxide (CO$_2$), water steam (H$_2$O) and nitrogen (N$_2$)~\cite{bulinska2017cfd}, \done{where each gas species (O$_2$, CO$_2$, H$_2$O, N$_2$) is assigned a constant density}. The numerical model is based on the Navier-Stokes equations for incompressible flow, incorporating continuity, momentum, energy conservation, and turbulence model equations, along with species transport equations.

\done{The general form of the continuity equation is:
\begin{equation}
\frac{\partial \rho}{\partial t} + \nabla \cdot (\rho \bm{u}) = 0,
\end{equation}
where $\rho$ is fluid density and $\bm{u}$ is the velocity field. In our simulations, ANSYS Fluent solves this equation under the incompressible flow assumption by setting the density $\rho$ as constant. 
}
The conservation equations for air constituents govern the transport and distribution of individual gas species within the airflow.
\begin{equation}
    \frac{\partial Y_i}{\partial t} + \nabla \cdot (Y_i \bm{u}) = -\nabla \cdot \frac{1}{\rho}\bm{j}_i, 
\end{equation}
where $i$ denotes three air constituents, namely, O$_2$, CO$_2$ and H$_2$O, $Y_i$ is the mass fraction of the $i$-th air constituent, and $\rho$ is density of fluid.
The mass flux of the $i$-th constituent can be calculated $\bm{j}_i = -D_{\text{eff}}\nabla Y_i$,
where $D_{\text{eff}}$ is the effective diffusion coefficient which includes turbulence effects. The mass fraction of N$_2$ was calculated from the sum of mass fractions of all air species which should be equal to unity. CO$_2$ concentration can be converted from mass fraction of CO$_2$ with
\begin{equation}
    C(\x,t) = Y_{\text{CO$_2$}}(\x,t) \cdot 10^6 \cdot \frac{\text{molecular weight of CO$_2$}}{\text{molecular weight of air}},
\end{equation}
where molecular weight of CO$_2$ = $44.01\text{g/mol}$ and molecular weight of air is ${28.97\text{g/mol}}$. 
The momentum equation describes the motion of air as an incompressible fluid, governed by the Navier-Stokes equations:
\begin{equation}
    \frac{\partial (\rho \bm{u})}{\partial t} + \nabla \cdot (\rho\bm{u}\bm{u}) = -\nabla p + \rho \bm{g}  + \nabla \cdot (\mu \nabla \bm{u}) - \nabla \cdot \tau_t, 
\end{equation}
where $p$ is the pressure field, $\bm{g}$ is a vector of gravitational acceleration, $\mu$ is a molecular dynamic viscosity and $\tau_t$ is a turbulence tensor. 
The energy conservation equation governs the transport of thermal energy within the airflow,
\begin{equation}\label{eq:energy_equation}
   \done{ \frac{\partial (\rho e)}{\partial t} + \nabla \cdot (\rho e\bm{u}) = \nabla \cdot (k_{\text{eff}}\nabla T) - \nabla \left(\sum_i h_i \bm{j}_i\right), }
\end{equation}
where $e$ is a specific internal energy, $k_{\text{eff}}$ is an effective heat conductivity, $T$ is fluid temperature and $h_i$ refers to a specific enthalpy of fluid. 
The turbulence model equations approximate the effects of small-scale turbulent eddies 
$$\tau_{t,ij} = \mu_t \left(\frac{\partial \bm{u}_i}{\partial x_j} + \frac{\partial \bm{u}_j}{\partial x_i}\right) - \frac{2}{3}\rho \kappa \delta_{ij}\,,$$
where $\mu_t$ is a turbulent viscosity, $\kappa$ is a turbulent kinetic energy and $\delta_{ij}$ is Kronecker's delta. 

The domain boundary $\partial Z$ encompasses all surfaces, including walls, ventilation interfaces, and occupant boundaries. We further define $\Z_{\text{supply}}$ as the inlet vent boundary, $\Z_{\text{occupant}}$ as the occupant surface boundary, $\Z_{\text{return}}$ as the outlet vent boundary. 
The airflow boundary conditions are then given by:
\begin{subequations}
\begin{align}
  &  \bm{u}(\x,t) = m^r_i(t) \begin{bmatrix}
        \sin(m^a_{i}(t)) \\ 0 \\  -\cos(m^a_{i}(t))
    \end{bmatrix}, 
    \forall \x \in \Z_{\text{supply},i}, \label{eq:air_c1} \\
& \bm{u}(\x,t) = \frac{v_{\text{occupant}}}{A_{\text{occupant}}} \cdot  N_{\text{occupant}} \cdot \begin{bmatrix}
    0 \\ 0 \\ 1
\end{bmatrix}, \forall \x \in \Z_{\text{occupant}}, \label{eq:air_c4}\\
   &  \bm{n} \cdot \nabla \bm{u} = 0, \forall z \in \mathcal{Z}_{\text{return}}, \label{eq:air_c2}  \\
   & \bm{u}(\x,t) = 0, \forall \x \in \partial \Z  \setminus(\Z_{\text{supply}} \cup \Z_{\text{occupant}} \cup \Z_{\text{return}}). \label{eq:air_c3}
\end{align}
\end{subequations}
\eqref{eq:air_c1} specifies that the airflow velocity at $i$-th group of supply vents is determined by its corresponding airflow rate $m^r_i(t)$ and direction angle $m^a_i(t)$. \eqref{eq:air_c4} relates the airflow rate to the number of occupants, where the exhaled air rate $v_{\text{occupant}}$ is set to 6 L/min per person~\cite{he2022experimental}, and $A_{\text{occupant}}$ denotes the occupant boundary area. 
Constraint~\eqref{eq:air_c2} sets the Neumann boundary conditions at the return vent and constraint~\eqref{eq:air_c3} applies Dirichlet conditions to all other boundaries by setting the airflow velocity as zero~\cite{bian2024ventilation}. 


\color{donecolor}
\section{Learning results of deep learning-based reduced-order modeling of CFD data}
\label{ap:DL-ROM}

We compare the proposed neural operator learning approach against a state-of-the-art deep learning-based reduced-order modeling for predicting the spatial-temporal dynamics of CO$_2$ concentration and building ventilation control. 
Specifically, follow the work~\cite{pant2021deep}, which employed 3D autoencoder and U-Net architectures to perform nonlinear reduced-order modeling of CFD dynamics. We then integrate this learned surrogate model into our ventilation control optimization~\eqref{eq:problem} to optimize control actions. 

\begin{figure*}[htbp]
    \centering
    \includegraphics[width=1.0\linewidth]{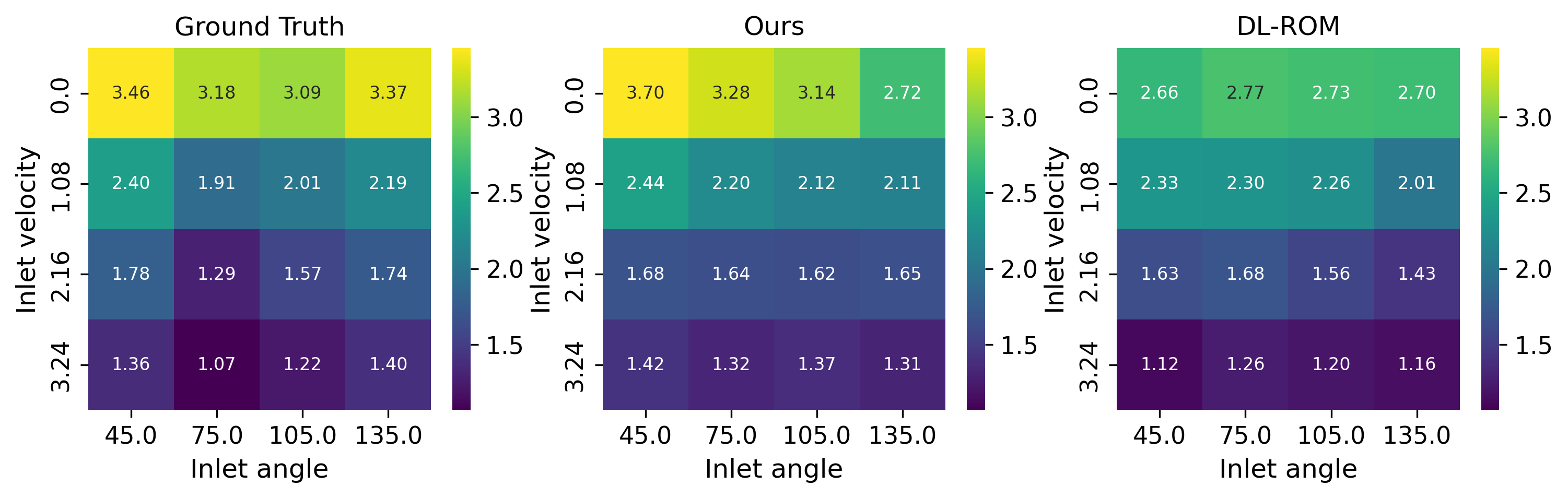}
    \caption{Visualization of control objective values $\|\widehat{C} - C_{\text{target}}\|_2^2$ across a grid of control inputs (airflow rate and vent angle), evaluated using (left) ground truth CFD simulations, (middle) our neural operator ensemble, and (right) DL-ROM~\cite{pant2021deep}. }
    \label{fig:learn_control}
\end{figure*}

Apart from $l_2$ error, we propose a new metric, Control Landscape Error (CLE), to quantify discrepancies between ground-truth CFD results and model predictions (our neural operator or DL-ROM~\cite{pant2021deep}) under different control conditions. We generate 16 unique control combinations by uniformly sampling airflow angles from $45^\circ$ to $135^\circ$ and rates from 0 to 3.24\,m/s, applied identically across all six vents. For each, we simulate the future CO$_2$ distribution and use it to evaluate model accuracy across varying control inputs.
We include a representative visualization in Figure~\ref{fig:learn_control} to complement this quantitative evaluation.
The control objective includes the term $\|\widehat{C} - C_{\text{target}}\|_2^2$, which measures the deviation of the predicted CO$_2$ concentration from the target value. 
We visualize this objective value (with CO$_2$ concentrations standardized using Z-score normalization) for the ground truth CFD simulations, our neural operator ensemble, and the DL-ROM baseline. Our model captures the control-performance landscape more accurately, especially in lower-velocity regimes (that saves energy), whereas DL-ROM tends to over-smooth the variations, particularly at the extremes of the action space. This result demonstrates that our neural operator preserves the input-output sensitivity of the system, which is crucial for robust control optimization.

The model performance is reported in Table~\ref{tab:learn_control}, where CLE is defined as  
\begin{equation}
\text{CLE} = \frac{1}{N}\sum_{i=1}^N \left| \|C_{i,\text{gt}} - C_{\text{target}}\|_2^2 - \|C_{i,\text{model}} - C_{\text{target}}\|_2^2 \right|,
\end{equation}  
where $ C_{i,\text{gt}} $ and $ C_{i,\text{model}} $ denote the ground truth and model-predicted CO$_2$ fields for the $i$-th control input, and $N$ is the number of evaluated inputs. Here, we generate 2 test samples (requiring approximately 32 CPU hours in total), each containing 16 control input pairs, resulting in $N = 32$ evaluations. While DL-ROM achieves a slightly lower $l_2$ error on the test set, our operator-based model attains a lower CLE, indicating better accuracy in capturing the control-response relationship. This highlights a key strength of our method: although the global $l_2$ error is comparable, our model more faithfully preserves the sensitivity of CO$_2$ outcomes to varying control inputs.

\begin{table}[H]
    \centering
    \begin{tabular}{lp{1.5cm}p{1.5cm}p{2cm}} \hline 
     Model    &   $l_2$ error (train dataset) & $l_2$ error (test dataset) & Control Landscape Error(CLE)    \\ \hline 
     Ours    &  \textbf{5.9\%}  &  10.9\%      &  \textbf{0.17}    \\ 
     DL-ROM  &  8.4\%   &  \textbf{9.6\%}  &  0.26    \\ \hline
    \end{tabular}
    \caption{$l_2$ error and Control Landscape Error (CLE) for each model, with the best performance highlighted in bold. }
    \label{tab:learn_control}
\end{table}